\documentclass{jfm}
\usepackage{graphicx} \usepackage{natbib}
\usepackage{upmath,amsmath,amssymb,amsbsy}

\providecommand\boldsymbol[1]{\mbox{\boldmath $##1$}}

\newcommand\Rey{\mbox{\textit{Re}}}  
\renewcommand{\v}{\ensuremath{\mathrm{{\mathbf u}}}} 
\newcommand\ci{\mathrm{i}}

\newsavebox{\astrutbox}
\sbox{\astrutbox}{\rule[-5pt]{0pt}{20pt}}

\newcommand\eg{e.g.\ }
\newcommand\ie{i.e.\ }

\graphicspath{{./fig/}}

\title[On the transient nature of localized pipe flow
  turbulence]{On the transient nature of localized pipe flow
  turbulence}
\author[M. Avila, A. P. Willis and B. Hof]{M\ls A\ls R\ls C\ns A\ls
  V\ls I\ls L\ls A\ns$^1$,\ns A\ls S\ls H\ls L\ls E\ls Y\ns P.\ns W\ls
  I\ls L\ls L\ls I\ls S$^2$ \and B\ls J\ls \"O\ls R\ls N\ns H\ls O\ls
  F$^1$}

\affiliation{$^1$Max Planck Institute for Dynamics and
  Self-Organization, 37073 G\"ottingen, Germany\\[\affilskip]
  $^2$Laboratoire d'Hydrodynamique, \'Ecole Polytechnique, 91128
  Palaiseau, France}

\begin{document}

\maketitle

\begin{abstract}

The onset of shear flow turbulence is characterized by turbulent
patches bounded by regions of laminar flow. At low Reynolds numbers
localized turbulence relaminarises, raising the question of whether it
is transient in nature or it becomes sustained at a critical
threshold. We present extensive numerical simulations and a detailed
statistical analysis of the data, in order to shed light on the
sources of the discrepancies present in the literature. The results
are in excellent quantitative agreement with recent experiments and
show that the turbulent lifetimes increase super-exponentially with
Reynolds number. In addition, we provide evidence for a lower bound
below which there are no meta-stable characteristics of the
transients, \ie the relaminarisation process is no longer memoryless.

\end{abstract}

\section{Introduction}

The development of turbulence in shear flows poses a challenge of
great theoretical and practical relevance
\citep{grossmann2000,eckhardt2007}. Since the seminal work of
\citet{reynolds1883} on the onset of turbulent fluid motion in a
circular pipe, this system has remained a paradigm for transition
without linear instability, \ie subcritical transition. Here, the
boundary that separates the laminar flow from turbulence depends not
only on the Reynolds number (\Rey), but also on the characteristics of
ambient and external perturbations. In particular, the threshold in
perturbation amplitude that must be exceeded to trigger transition
scales as $\Rey^{-\gamma}$, with $\gamma\in[1,2]$ depending on the
perturbation details
\citep*{hof2003,peixinho2007,mellibovsky2009}. Thus, if great care is
taken to minimize all sources of disturbances, the flow can be kept
laminar up to \Rey\ as large as $10^5$ \citep{pfenniger1961}.

\citet{brosa1989} showed that even long times after turbulence is
initially triggered, relaminarisation to Hagen-Poiseuille flow may
occur. \citet{faisst2004} systematically studied the probability of a
turbulent trajectory surviving up to time $t$, given by the survivor
function
\begin{equation}\label{eq:survivor}
  S(t)=P(\mbox{flow is turbulent at } T \ge t).
\end{equation}
They concluded that relaminaristaion occurs suddenly and that the
process is memoryless, \ie lifetimes are exponentially distributed
with $S(t)=\exp[t/\tau]$, where $\tau$ is the mean turbulent
lifetime. This behavior had been previously observed in turbulent
relaminarisation experiments in plane Couette flow
\citep{bottin1998a,bottin1998b}. The exponential distribution of
lifetimes and sensitive dependence on initial conditions
\citep*{darbyshire1995,faisst2004} is consistent with the presence of
a repellor in phase space, a strange saddle \citep[see
  e.g.][]{kerswell2005,eckhardt2008}. The skeleton of the saddle is
then constructed from exact unstable solutions of the Navier--Stokes
equations, and the simplest of such solutions correspond to nonlinear
travelling waves, discovered numerically by \citet{faisst2003} and
\citet{wedin2004}.  Close visits to such travelling waves in turbulent
flows were reported in the experiments of \citet{hof2004} and
simulations by \citet{kerswell07}.  Recently, new classes of
travelling waves solutions and relative periodic orbits have also been
found \citep*{pringle2007,duguet2008}.

\citet{faisst2004} investigated the scaling of mean turbulent
lifetimes as \Rey\ is increased and proposed a divergence of
reciprocal form $\tau(\Rey) \propto (\Rey_c-\Rey)^{-1}$, indicating
that turbulence becomes self-sustained at $\Rey_c \simeq 2250$.
However, a limitation of the lifetime study of \citet{faisst2004} was
the length of the periodic pipe used in their simulations ($5D$),
chosen to reduce computational costs. At $\Rey \lesssim 2400$ the
experimentally observed flow regimes consist of localized turbulent
structures bounded by regions of laminar flow, so-called {\it puffs}
\citep{wygnanski1973}.  Puffs have a length of about $20D$ and are
characterized by a sharp turbulent-laminar interface at the trailing
edge and a slowly diffused leading edge (see
figure~\ref{fig:puff}). Thus, long periodic pipes are required to
fully capture their relevant dynamics. The first experimental study to
determine lifetimes in pipe flow was performed by
\citet{peixinho2006}, who carried out \Rey\ {\it reduction}
experiments following the methodology introduced by
\citet{bottin1998b} in plane Couette flow. Using a constant-mass-flux
pipe of $750D$, they perturbed the laminar profile to generate a puff,
then reduced \Rey\ and measured the downstream distance at which the
puff decayed. Their study also supported a divergence of lifetimes of
reciprocal form at $\Rey_c=1750$.  An experimental study performed in
a gravity-driven pipe by \cite{hof2006}, however, rendered an
exponential scaling of $\tau$ with \Rey, indicating that turbulence
may be transient for all finite \Rey.

\citet{willis2007} performed numerical simulations of a periodic
constant-mass-flux $50D$ pipe, which allowed for direct comparison
with experiment for the first time. Their results qualitatively
supported the critical behavior observed by \citet{peixinho2006},
although the extrapolated critical Reynolds number $\Rey_c=1870$ was
larger than the experimentally measured value.  More recently,
\citet{hof2008} repeated the experimental study with substantial
technical improvements over the setup of \cite{hof2006}. The
implementation of accurate temperature control and automatisation of
the measurement techniques allowed them to prolong the observational
time-span up to $8$ orders of magnitude, drastically extending all
previous investigations. Their measurements up to $\Rey=2050$ still
exhibited relaminarisation and indicated a faster than exponential
(but non-diverging) scaling of lifetimes with \Rey, providing evidence
that puff turbulence is of transient nature. While the data of
\citet{willis2007} was incompatible with an exponential scaling, good
agreement with the \emph{super}-exponential scaling of \citet{hof2008}
calls for further numerical investigation.

The main goal of this paper is to provide a comprehensive discussion
on the scaling of turbulent lifetimes in pipe-flows, reconciling
previous investigations and clarifying the sources of discrepancy. To
this aim, the numerical simulations of \cite{willis2007} have been
extended in \Rey, observation times and sample sizes. This has
substantially increased the statistical significance, rendering much
reduced and accurate confidence intervals. A detailed account of the
methodology and analysis of the data is presented and the influence of
initial conditions is examined. The results are in close quantitative
agreement with the recent experiments of \cite{hof2008} in
relaminarisation probabilities at each $\Rey$ investigated, and hence
support a super-exponential increase of the lifetimes with $\Rey$. In
addition, it is found that the process is not memoryless for $\Rey
\lesssim 1720$, suggesting a bifurcation event that gives rise to
turbulent trajectories.

\section{Numerical Method}\label{sec:methods}

Consider an incompressible viscous fluid which is driven through a
pipe of circular cross-section at a constant flow rate. The Reynolds
number is defined as $\Rey=UD/\nu$, where $U$ is the mean flow-speed,
$D$ the pipe diameter and $\nu$ the kinematic viscosity of the fluid.
It is convenient to scale lengths by $D/2$ and velocities by $2U$ in
the Navier-Stokes equations, leading to
\begin{equation}\label{eq:NSE}
  \partial_t\v + (\v\cdot\nabla)\v +\nabla p =
  \frac{1}{\Rey}\Delta\v,\quad \nabla\cdot\v = 0.
\end{equation}
These are supplied with no-slip boundary conditions at the pipe wall,
whereas in the axial direction the pipe is periodic.

The Navier--Stokes equations \eqref{eq:NSE} are solved in cylindrical
coordinates $(r,\theta,z)$ using the hybrid spectral finite-difference
method.  This is based on the velocity-potential
formulation of \citet{marques1990}, where the difficulties arising
from the coupled boundary conditions on the potentials are resolved
with the influence-matrix method
\citep{willis2009}.  Variables are expanded in Fourier modes
\begin{equation}\label{eq:specexp}
  A = \sum_{k=-K}^K\sum_{m=-M}^M A_{k,m}(r) \exp[\ci(\alpha k z + m \theta)].
\end{equation}
As the variables are real, their Fourier coefficients satisfy the
property $A_{km}=A_{-k,-m}^\dagger$, where $^\dagger$ denotes the
complex conjugate. In the results presented here $K=384$ and $M=24$,
corresponding to $\pm 384$ axial and $\pm 24$ azimuthal Fourier modes.
In the axial direction the wavelength of the pipe is $16 \pi D \simeq
50D$. This is sufficient to avoid the interaction between leading and
trailing edge of the localized structure of a puff (of approximately
$20D$) which would arise in short domains. In the radial direction the
explicit finite-difference method has been used on $9$-point stencils
in a grid of $N=40$ points.  The time-step was dynamically controlled
using information from a predictor-corrector method and was limited to
$\Delta t < 0.005D/U$.  A resolution test on lifetime statistics is
shown in the following section. For further details on the
formulation, methods and tests see \citet{willis2009}.

\section{Initial conditions}\label{sec:ic}

\begin{figure}
  \centering
  \includegraphics[width=0.98\linewidth]{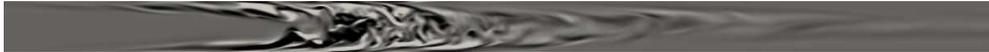}
  \caption{Instantaneous stream-wise vorticity distribution in a
    $(r,z)$-plane of a turbulent puff at $\Rey=2000$. In the axial
    direction $20D$ are shown from a periodic domain of $50D$.}
  \label{fig:puff}
\end{figure}

In order to generate the initial conditions for the lifetime
simulations, a localized disturbance was applied to the laminar
Poiseuille flow at $\Rey=2000$. The disturbance quickly evolved into
an `equilibrium puff' which remained constant in spatial extent while
propagating downstream (see figure~\ref{fig:puff}). The puff was
evolved for $t \sim 5000D/U$ and snapshots of the full velocity field
were recorded every $10D/U$, generating a collection of initial
conditions. Subsequently, runs at lower \Rey\ were performed starting
from these initial conditions and were monitored until the flow
relaminarised. The criterion for relaminarisation was that the energy
of the axially dependent modes drop below $0.005U^2D^3$, at which
point turbulent motions had decayed beyond recovery.

\begin{figure}
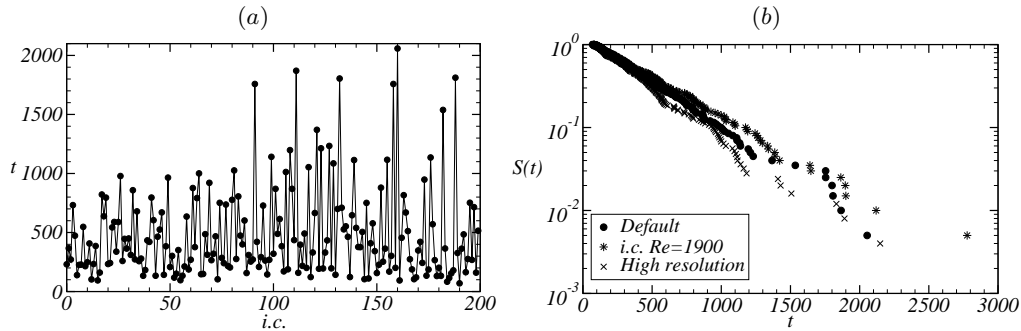

  \begin{center}
    \begin{tabular}{cc}
      $(a)$ & $(b)$ \\  
      \includegraphics[width=0.48\linewidth,clip=]{init_cond.eps} &
      \includegraphics[width=0.5\linewidth,clip=]{S_1860b.eps}
    \end{tabular}
  \end{center}
  \caption{$(a)$ Relaminarisation times for reduction simulations from
    $\Rey=2000$ to $\Rey=1860$ as a function of initial condition. $(b)$
    Survivor function at $\Rey=1860$ from the data in $(a)$ (circles),
    initial conditions at $\Rey=1900$ (stars) and high numerical
    resolution (crosses).}
  \label{fig:num}
\end{figure}

Figure~\ref{fig:num}$(a)$ shows relaminarisation times following a
reduction of the Reynolds number to $\Rey=1860$ as a function of
initial condition. The times from here onwards are shown in $D/U$
units. The large fluctuations in the data indicate that consecutive
initial conditions result in uncorrelated lifetimes. From these data,
the corresponding survivor function $S(t)$ has been obtained and is
plotted as circles in the logarithmic scale of
figure~\ref{fig:num}$(b)$. The approximately constant slope of $S(t)$
implies the memoryless nature of the relaminarisation process. To
verify the numerical resolution, the number of modes in the azimuthal
and axial directions was increased by $33\%$ and the number of
finite-difference points by $25\%$. The number of degrees of freedom
is greater than $2.2$ times that of the default resolution.  The
results are plotted in figure~\ref{fig:num}$(b)$ as crosses and are
within statistical certainty of the default prediction (see inset of
figure~\ref{fig:scaling}). Thus, we conclude that the turbulent puffs
are accurately resolved.

The impact of the mechanism to initiate the turbulent state on the
relaminarisation probabilities has been object of much debate in the
literature \citep{willis2007,schneider2008,delozar2009}. Indeed, the
reduction procedure of \citet{peixinho2006} was introduced as an
attempt to minimise the effect of initial conditions in lifetime
measurements. Robustness of the survivor function has been examined
here by repeating the lifetime simulations at $\Rey=1860$ but using
three additional sets of initial conditions, each of them consisting
of $200$ velocity field snapshots. First, initial conditions of a puff
simulation at $\Rey=2200$ were used, representing a much greater
initial change of the $\Rey$ over the default reduction from
$\Rey=2000$. This was again repeated but from a puff simulation at
$\Rey=1925$. Finally, following \citet{willis2007} randomly selected
velocity snapshots of several puffs at $\Rey=1900$ were used. The
stars in figure~\ref{fig:num}$(b)$ show the results of the reduction
from $\Rey=1900$, whereas results from $\Rey=2200$ and $1925$ are not
shown to avoid crowding the figure. All three additional cases render
characteristic lifetimes within the $95\%$ confidence interval about
the default prediction (see inset of figure~\ref{fig:scaling}),
higlighting the independence of the lifetime statistics from initial
conditions. These results confirm the experimental findings of
\citet{delozar2009}, who have demonstrated the invariance of the
lifetime distributions by using four different experimental protocols
to generate the turbulent puffs.

\section{Statistical analysis of turbulent lifetimes}\label{sec:stat}

An exponential distribution of turbulent lifetimes is one of the main
pillars of the strange saddle paradigm for the transition to
turbulence in shear flows \citep{eckhardt2008}. In practice, the
distributions are exponential only for $t>t_0>0$, where $t_0$ depends
on the procedure to initiate turbulence and Reynolds number in a
non-trivial manner \citep{schneider2008,delozar2009}. When \Rey\ is
suddenly quenched in the reduction experiments, the puff is forced to
quickly readjust to the new flow rate and this could cause immediate
relaminarisation. In phase space, this is equivalent to a trajectory
not having approached the strange saddle. Thus, consistently
determining $t_0$ constitutes a challenge in lifetime studies of shear
flows. The problem stems from the difficulties in discriminating
between a trajectory that has shown very short turbulent behavior and
one that has directly relaminarised. This is especially delicate in the
low \Rey\ regime, where a large part of the sample features lifetimes
shorter than $t_0$.

The hazard function of a lifetime distribution,
defined as
\begin{equation}\label{eq:hazard}
  h(t)=-\dfrac{d}{dt}\log(S(t)),
\end{equation}
specifies the instantaneous rate of death at time $t$, \ie $h(t)\delta
t$ is the approximate probability of death in $[t,t+\delta t]$ given
survival up to time $t$. For an exponential distribution, the hazard
function is constant $h(t)=1/\tau$, where $\tau=\langle t\rangle$ is
the mean lifetime of the population. This corresponds to a memoryless
process. In the theory of dynamical systems, the escape rate from the
saddle is defined as $\kappa=1/\tau$ \citep{tel2008}.

\subsection{Determining $t_0$: onset of the strange saddle}\label{sec:saddle}

\begin{figure}
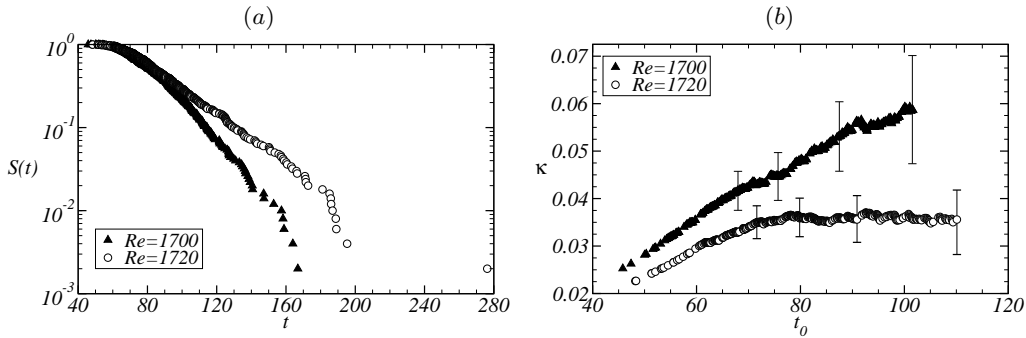

  \begin{center}
    \begin{tabular}{cc}
      $(a)$ & $(b)$ \\  
      \raisebox{3pt}{
        \includegraphics[scale=0.256,clip=]{S_1720.eps}
      } &
      \includegraphics[scale=0.256,clip=]{exp_1720.eps}
    \end{tabular}
  \end{center}
  \caption{$(a)$ Survivor function of turbulent lifetimes at
    $\Rey=1720$ (circles) and $1700$ (triangles) obtained with $500$
    simulations each. $(b)$ Escape rates $\kappa=1/\tau$ as a function
    of $t_0$.  The error bars correspond to $95\%$ confidence
    intervals around the $\kappa$ predictions. }
  \label{fig:re1720}
\end{figure}

Figure~\ref{fig:re1720}$(a)$ is a logarithmic plot of the lifetime
distribution at $\Rey=1700$ (triangles) and $\Rey=1720$ (circles).
Their corresponding escape rates have been computed by subsequently
excluding from the sample puffs that have decayed before a given time
$t_0$, \ie $\kappa(t_0)=1/\tau(t_0)$ with $\tau(t_0)=\langle t-t_0 |
t>t_0 \rangle$. Figure~\ref{fig:re1720}$(b)$ shows the resulting
values of $\kappa(t_0)$ as a function of cutoff time $t_0$. Note that
as $t_0$ is increased, the size of the sample used to estimate
$\kappa$ is progressively reduced from $500$ to $100$, resulting in
greater error bars. For $\Rey=1720$, the plot approaches a horizontal
line at about $t_0\simeq72$, indicating a constant hazard function and
thus exponentially distributed lifetimes. Hence, we conclude that at
$\Rey=1720$ the escape rate of turbulent trajectories from the saddle
is $\kappa\simeq 0.035$, whereas trajectories that have decayed before
$t_0\simeq 72$ cannot be counted as having visited the saddle. Overall
$t_0$ increases slightly from $72$ at $\Rey=1720$ to $95$ at
$\Rey=1820$.  Meanwhile, the number of trajectories decaying without
visiting the saddle decreases as \Rey\ increases, and at $\Rey=1860$
almost all trajectories are attracted to the saddle.

No evidence of exponential behaviour in the lifetimes for
$\Rey<1720$ has been found.  This is illustrated by the hazard
function at $\Rey=1700$, plotted as triangles in
figure~\ref{fig:re1720}$(b)$. Despite the initial similarity with
$\Rey=1720$, the escape rate for $\Rey=1700$ does not settle to a
constant value but increases monotonously.  This implies that the
probabilities of relaminarisation depend on the history of the
turbulent trajectory.  Similar results have been found for
$\Rey=1580$ and $1640$ using $500$ simulations in each case. Hence, the
data at low $\Rey<1720$ cannot be used to extrapolate lifetime scaling
with $\Rey$, at least not under the assumption of a memoryless process
corresponding to the escape from a strange saddle.

\subsection{Estimation of characteristic turbulent lifetimes}\label{sec:samp}

Characteristic turbulent lifetimes are estimated from an exponential
distribution,
\begin{equation}
  S(t)=\exp[(t-t_0)/\tau_{\text{true}}],
\end{equation}
with the sample mean, which is the the Maximum Likelihood Estimator
(MLE) of the parameter $\tau_{\text{true}}$. However, the observation
time-span is finite in practice, which implies that the data need to
be truncated and therefore the sample mean cannot be obtained. A
lifetime sample of size $n$ with truncation after $r$ decays is known
as censored data of Type-II \citep{lawless2003}. In
this case, the MLE of $\tau_{\text{true}}$ is given by
\begin{equation}\label{eq:mle}
  \tau=\dfrac{1}{r}\Big[\sum_{i=1}^r t_i + (n-r)t_r\Big],
\end{equation}
where $t_0+t_i$ is the lifetime of the $i$th puff to decay, hence
$t^*=t_0+t_r$ being the time where the simulations were truncated. The
corresponding exact confidence intervals, at level $1-\alpha$, are
\begin{equation}\label{eq:conf}
  \tau_{\text{true}} \, \in \,
  \tau\times\big[2r/\chi^2_{2r,1-\alpha/2},2r/\chi^2_{2r,\alpha/2}\big],
\end{equation}
where $\chi^2_{m,p}$ is the $p$th quantile of the chi-squared
distribution with $m$ degrees of freedom. It is worth noting that for
uncensored data ($r=n$), the sample mean is recovered as MLE, and more
importantly, that the relative size of the confidence intervals is
uniquely determined by the number of runs $r$ that have decayed. In
the uncensored case the Central Limit Theorem yields approximate
$95\%$ confidence intervals $\tau\times(1\pm1.96/\sqrt n)$, and the
need for a large number of observations is a consequence of the slow
convergence with $n$.

\begin{figure}
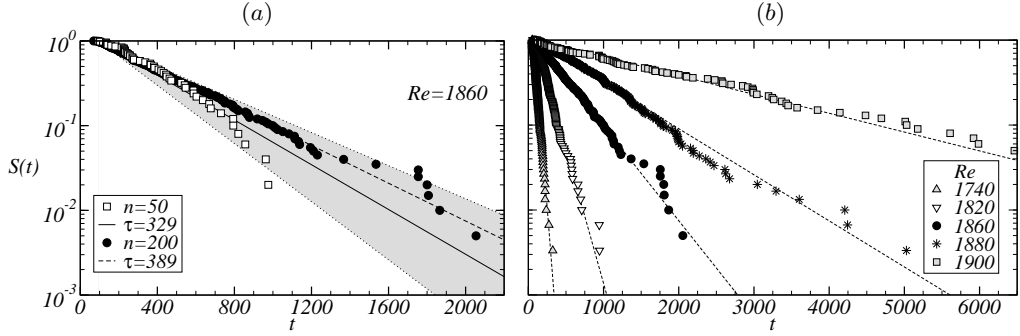

  \begin{center}
    \begin{tabular}{cc}
      $(a)$ & $(b)$ \\  
      \includegraphics[width=0.49\linewidth,clip=]{S_1860s.eps}&
      \includegraphics[width=0.49\linewidth,clip=]{prob_all.eps}
    \end{tabular}
  \end{center}
  \caption{$(a)$ Survivor function at $\Rey=1860$ obtained from sample
    sizes $n=50$ (squares) and $200$ (circles). The solid (dashed)
    line is the curve $\exp[(t-t_0)/\tau]$ for $n=50$ ($200$), where
    $\tau$ has been estimated from \eqref{eq:mle}. The shaded area
    between the two dotted lines corresponds to the confidence
    interval \eqref{eq:conf} for $n=50$. $(b)$ Survivor function of
    turbulent lifetimes at several Reynolds numbers. For clarity, only
    part of the \Rey\ investigated are shown.}
  \label{fig:proball}
\end{figure}

Figure~\ref{fig:proball}$(a)$ shows the survivor function at
$\Rey=1860$ for the first $50$ runs in figure~\ref{fig:num}$(a)$ and
for the full sample $n=200$. The mean lifetimes have been estimated
from \eqref{eq:mle} and are plotted as solid and dash lines
illustrating the curves $\exp[(t-t_0)/\tau]$. Although the
characteristic lifetime from the sub-sample $n=50$ is shorter, its
$95\%$ confidence interval \eqref{eq:conf}, spanned by the shaded area
in the figure, includes the result from the full sample.

\section{Lifetime scaling with \Rey: the transient nature of puff turbulence}\label{sec:scale}

In order to shed light on the discrepancies in scaling of turbulent
lifetimes present in the literature, we have extended the Reynolds
number range in the numerical simulations up to $\Rey=1900$. The
results are presented in figure~\ref{fig:proball}$(b)$, showing
survivor functions at several $\Rey$ in a logarithmic scale. The
sample sizes have been subsequently increased until each data-set
unambiguously showed the exponential distribution. In particular,
between $200$ and $500$ simulations were run for $\Rey\in[1720,1880]$,
whereas at $Re=1900$ only $100$ cases were simulated due to the
computational costs incurred in very long time integrations. In
addition, the simulations at $Re=1900$ were stopped when $96$ puffs
had decayed, leaving only $4$ survivors at $t^*\simeq 6500$.

\begin{figure}
  \begin{center}
    \includegraphics[width=0.98\linewidth,clip=]{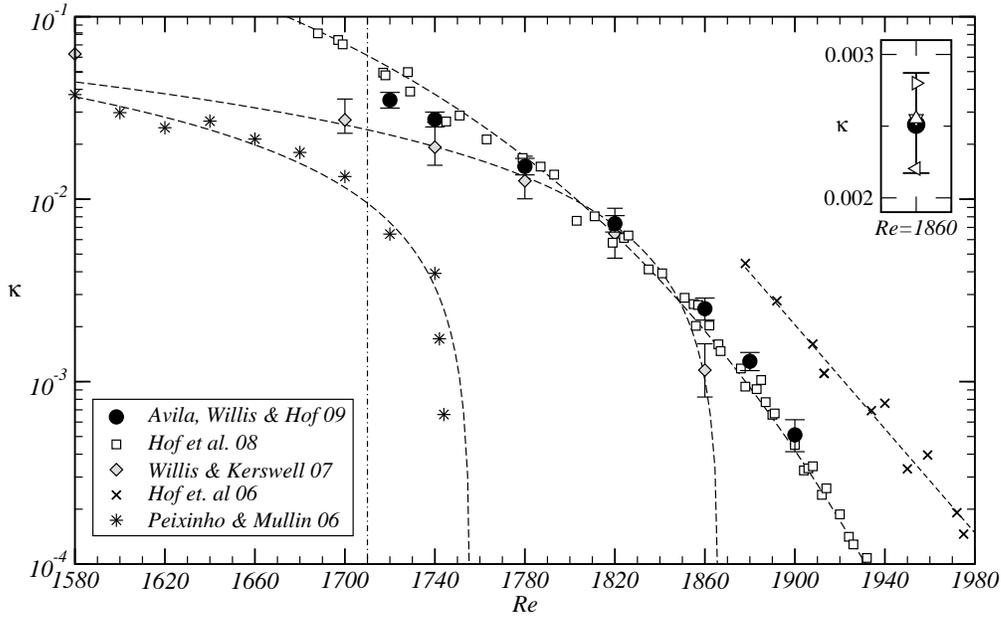}
  \end{center}
  \caption{Escape rate $\kappa$ scaling with $\Rey$ as obtained from
    the mean lifetimes of the survivor functions in
    figure~\ref{fig:proball} (shown in black circles). The vertical
    line marks the approximate onset of the strange saddle. The inset
    shows the escape rate at $\Rey=1860$: default reduction from
    $\Rey=2000$ (black circle and $95\%$ confidence interval),
    reductions from $\Rey=2200,1925,1900$ ({\scriptsize
      $\triangle$},$\triangledown$,$\triangleleft$) and high numerical
    resolution ($\triangleright$).}
  \label{fig:scaling}
\end{figure}

From the survivor functions in figure~\ref{fig:proball}$(b)$ escape
rates $\kappa=1/\tau$ have been estimated using \eqref{eq:mle} and are
shown as black circles in the logarithmic scale of
figure~\ref{fig:scaling}. Confidence intervals at the $95\%$ level,
obtained from \eqref{eq:conf}, have been plotted and may be regarded
as error bars due to sample size. The results of the high resolution
runs and reductions from $\Rey=2200$, $1925$ and $1900$ to $\Rey=1860$
fall within the $95\%$ confidence interval of the default case and are
only shown in the inset to avoid overlapping the figure. Overall, the
results are in excellent quantitative agreement with the experimental
relaminarisation probabilities of \citet{hof2008}, shown as
squares. With increasing \Rey, $\kappa$ decreases extremely fast,
corresponding to a super-exponential increase in turbulent
lifetimes. It is worth noting that the measurements of \citet{hof2008}
extend up to $\Rey=2050$, with their super-exponential fit providing a
very good approximation over the full data set.

The agreement with the simulations of \citet{willis2007} is very good
for $\Rey=1740,1780,1820$. For $\Rey \lesssim 1720$ we have shown that
the lifetimes are not exponentially distributed (vertical line). The
discrepancy of the point at $\Rey=1860$ is attributed mainly to
statistical uncertainty. There $\tau_{\text{true}}$ was estimated with
$40$ simulations, of which only $28$ had decayed when truncated at
$t^*\simeq 1000$. The upper end of their confidence interval is close
to the experimental value of \citet{hof2008} and the present
computations, although there is still a small difference. We remark,
for example, that had the $40$ initial conditions $i\in[90,130]$ of
figure~\ref{fig:num}$(a)$ been used here to estimate
$\tau_{\text{true}}$, the result would be compatible with the estimate
provided by \citet{willis2007}. The escape rates and confidence
intervals shown in figure~\ref{fig:scaling} are those given in their
re-analysis \citep{willis2008arxiv}, using bootstrapping. Using the
method in \S\ref{sec:samp} on their original data shifts the point at
$\Rey=1860$ by only $10\%$ in the direction of the current data set,
with a $10\%$ larger confidence interval than that shown.

The experiments of \citet{hof2006} appear to be shifted by $2.5\%$ in
\Rey\ with respect to the numerical simulations and later experiments
of \citet{hof2008}. Their exponential fit, however, provides a good
approximation of the slope of $\kappa$ over $2$ orders of magnitude.
Similarly, applying a $7\%$ shift in \Rey, the data of
\citet{peixinho2006} overlaps for most of the range.  Significant
differences between their results and the results presented here, then
remain only for their two highest \Rey. A procedural difference is
the manner in which the puffs were generated. It has been shown here
and by \citet{delozar2009}, however, that the initial condition has no
influence other than on $t_0$. A more likely source of difference is
the difficulty of inferring the lifetime from the data. At high
$\Rey$, a short pipe implies very few relaminarisations and therefore
a very large range of possible $\tau$ according to (\ref{eq:mle}). For
example, at $\Rey=1745$ \citet{peixinho2006} obtained a mean lifetime
of $\tau=1515$. For an observation distance of $500D$ after reducing
\Rey, this corresponds to $r\le3$ decays out of $n=50$ experimental
runs, yielding a confidence interval
$\tau_{\text{true}}\times[629,7346]$. Nevertheless, we must
acknowledge that \citet{peixinho2006} were first to experimentally
demonstrate the very rapid increase in lifetimes over a small range of
Reynolds numbers.

\section{Conclusions}\label{sec:conc}

The transition to turbulence in pipe flow has been recently linked to
the presence of a strange saddle in the phase space of the
Navier--Stokes equations \citep[see \eg][]{eckhardt2008}. One of the
predictions of this emerging paradigm is that the turbulent lifetimes,
the time prior to escaping the saddle, should be exponentially
distributed. We have shown that the distributions are indeed
exponential by substantially increasing the sample size and thus
reducing statistical uncertainty with respect to previous works. In
order to recover the true escape rate, a substantially larger sample
size than previously used is necessary.

We have demonstrated that initial conditions from puffs at
$\Rey=2200,2000,1925,1900$ render the same characteristic lifetimes
after reduction to $\Rey=1860$, which verifies that the same turbulent
state is visited. Here, contrary to the possibility that the turbulent
state be difficult to reach when changing the Reynolds number, the
number of trajectories which fail to approach the saddle is actually
low (small $t_0$). This reflects structural similarity of the initial
condition with the state at the final $\Rey$.

Interestingly, the onset of the exponential distributions is rather
sharp in $\Rey$ at about $\Rey \lesssim 1720$.  This suggests a
bifurcation event that gives rise to the strange saddle. The onset of
meta-stable transients has been characterized in plane Couette flow
experiments by \citet{bottin1998a} and \citet{bottin1998b}, who
measured the average turbulent fraction. The change in distributional
shape revealed here provides an alternative reliable means of
quantifying a lower bound for meta-stable turbulence, and is
appropriate when considering single isolated disturbances.

Our results support the view of localised turbulence as a strange
repellor, with lifetimes increasing super-exponentially with \Rey\ as
in the experiments of \citet{hof2008}. A similar scaling has been
recently reported by \citet*{borrero2009} in Taylor--Couette flow with
stationary inner cylinder and previously by \citet{schoepe2004} in
super-fluid turbulence. To our knowledge the quantitative agreement of
the presented numerical simulations with experiment is unprecedented
in transition studies. It validates the use of periodic boundary
conditions and testifies the high demands on numerical resolution and
domain sizes which are required to faithfully capture the relevant
dynamics of turbulence.

\begin{acknowledgements} 
We wish to thank many cited authors for interesting and insightful
discussions regarding lifetimes in shear flows. M.~Avila and B.~Hof
are supported by the Max Planck Society. A.~P.~Willis is supported by
the E.C., Marie Curie Fellowship PIEF-GA-2008-219223.
\end{acknowledgements} 

\bibliographystyle{jfm}

\end{document}